\documentclass[aip,reprint]{revtex4-1}

\usepackage{amsmath,amssymb}
\usepackage[usenames]{color}
\usepackage{fancyhdr}
\usepackage[normalem]{ulem}

\usepackage{subfigure}
\usepackage{graphicx}
\usepackage{dcolumn}
\usepackage{bm}

\draft 

\begin{document}

\title{Short desynchronization episodes prevail in synchronous dynamics of human brain rhythms}

\author{Sungwoo Ahn}
\email{ahnmath@gmail.com}
\affiliation{Department of Mathematical Sciences and Center for Mathematical Biosciences, Indiana University Purdue University Indianapolis, IN 46032, USA}
\author{Leonid L. Rubchinsky}
\email{leo@math.iupui.edu}
\affiliation{Department of Mathematical Sciences and Center for Mathematical Biosciences, Indiana University Purdue University Indianapolis, IN 46032, USA}
\affiliation{Stark Neurosciences Research Institute, Indiana University School of Medicine, Indianapolis, IN 46032, USA }

\date{\today}

\begin{abstract}
Neural synchronization is believed to be critical
for many brain functions. It frequently exhibits temporal
variability, but it is not known if this variability has a specific
temporal patterning. This study explores these
synchronization/desynchronization patterns. We employ recently
developed techniques to analyze the fine temporal structure of
phase-locking to study the temporal patterning of synchrony of the
human brain rhythms. We study neural oscillations recorded by EEG in
$\alpha$ and $\beta$ frequency bands in healthy human subjects at
rest and during the execution of a task. While the phase-locking
strength depends on many factors, dynamics of synchrony has a very
specific temporal pattern: synchronous states are interrupted by
frequent, but short desynchronization episodes. The probability for
a desynchronization episode to occur decreased with its duration.
The transition matrix between  synchronized and desynchronized
states has eigenvalues close to 0 and 1 where eigenvalue 1 has
multiplicity 1, and therefore if the stationary distribution between
these states is perturbed, the system converges back to the
stationary distribution very fast. The qualitative similarity of
this patterning across different subjects, brain states and
electrode locations suggests that this may be a general type of
dynamics for the brain. Earlier studies indicate that not all
oscillatory networks have this kind of patterning of
synchronization/desynchronization dynamics. Thus the observed
prevalence of short (but potentially frequent) desynchronization
events (length of one cycle of oscillations) may have important
functional implications for the brain. Numerous short
desynchronizations (as opposed to infrequent, but long
desynchronizations) may allow for a quick and efficient formation
and break-up of functionally significant neuronal assemblies.
\end{abstract}

\pacs{87.19.lm 05.45.Xt}
\maketitle

\begin{quotation}
Neural synchrony is hypothesized to be important for
many physiological processes. Most of the time this synchrony is not
very strong, so that while neural signals may synchronous on the
average, they go in and out of phase. We found that neural
synchronization in human brain follows very specific temporal
pattern: synchronous states are interrupted by frequent, but short
desynchronization episodes. In general, the same synchrony strength
may result from many short desynchronization events or from few long
desynchronization events (as well as from a wide spectrum of
possibilities in between these extremes). However, in human brain
rhythms the probability for a desynchronization episode to occur
decreased with its duration. In addition, the transition matrix
between synchronized and desynchronized states has eigenvalues close
to 0 and 1 (the latter has multiplicity 1) promoting a very quick
convergence to a (presumably beneficial) stationary state after a
perturbation. The qualitative similarity of the fine temporal
structure of synchrony patters across different subjects, brain
states and parts of the brain suggests that this may be a general
type of dynamics for the brain. Earlier studies indicate that not
all oscillatory networks have this kind of patterning of
synchronization/desynchronization dynamics. Thus the observed
prevalence of short (but potentially frequent) desynchronization
events is likely to have important functional implications for the
brain. From a cell assembly theory view point the results may suggest that whenever a
cell assembly must be formed to facilitate a particular function or
task, short desynchronization dynamics may allow for a quick and
efficient formation and break-up of such an assembly.
\end{quotation}

\section{\label{intro}Introduction}
Neural synchrony is believed to be an important mechanism underlying
many phenomena in the human brain.~\cite{buzsaki_draguhn, uhlhaas,
fell_axmacher} It has been extensively studied using approaches and
methods of physics and nonlinear dynamics (see
Refs.~\onlinecite{rabinovich2006, nowotny2008}). Neural synchrony
strength is likely to be variable in time. The neural oscillations
are known to exhibit intermittent synchronization in both healthy
and diseased human and animal brain even at rest (see
Refs.~\onlinecite{velazquez, hurtado2005, gong2007,
hramov-koronovskii, park_rubchinsky1}). It was suggested that
transient dynamics in the nervous system is
generic.~\cite{rabinovich2008} Neural signals may go in and out of
synchrony due to variety of factors and approaches to detect and
quantify the presence of this variable and weak synchrony have been
considered (see Refs.~\onlinecite{hurtado_rubchinsky_2004,
levanquyen}). However, the properties of how this synchrony is
patterned in time have not being explored in normal human subjects.

Since this synchrony is not perfect, the same synchrony strength may
be achieved with markedly different temporal patterns of activity
(roughly speaking oscillations may go out of the phase-locked state
for many short episodes or few long episodes). However, synchrony is
a non-instantaneous phenomenon and from the data analysis
perspective one considers synchrony in a statistical sense, observed
over a sufficiently large number of cycles of
oscillations.~\cite{arkady_pikovsky}  Yet if this synchrony is
present on the average, one can look at each cycle of oscillations
and see how far away it is from a synchronized
state.~\cite{ahn_park_rubchinsky} This approach can describe the
differences in the dynamics and temporal structure of
synchronization/desynchronization events for the systems with
similar overall level of phase locking or similar stability of
synchronized state.~\cite{ahn_park_rubchinsky} This is especially
important given that the neural synchrony in the current study, and
a number of other neural systems, is not very strong. The underlying
network of presumably weakly coupled oscillators spends a
substantial fraction of time in the non-synchronous state.  Thus the
focus on desynchronization episodes is very reasonable.

 This approach has been recently applied to study the temporal patterns of
the pathological synchronization in subcortical brain areas in a
group of patients with Parkinson's disease.~\cite{park_rubchinsky1,
park_rubchinsky2} Locally measured synchronous (on the average)
oscillatory activity was observed to follow a specific pattern: the
synchronized state was interrupted by numerous but mostly short
desynchronization states. However it was not known if this was a
signature of Parkinson's disease, a feature of the specific
subcortical location, or a more general phenomenon. Here we show
that the latter is more likely to be the case. Since synchrony is
important to facilitate interactions between
neurons,~\cite{buzsaki_draguhn, uhlhaas, fell_axmacher} the temporal
patterning of this synchrony becomes a salient issue. We study
synchronized patterns of neural activity in a large sample of
healthy humans at rest and during an execution of a task. Similar
temporal patterning of synchronous activity in large cortical areas
in different states may suggest that i) this type of patterning is a
generic phenomenon in the brain, ii) it may have some functional
advantages for oscillating neural networks receiving, processing,
and transmitting information, iii) it may be grounded in some
general properties of neuronal networks calling for the development
of appropriate nonlinear dynamical theory.

\section{Experimental Data}
We used 64-channel electroencephalograms (EEG) of the international
10-10 system at the sampling rate of 160 Hz recorded from 109 normal
human subjects using the BCI2000 system~\cite{schalk_bci, bci2000}
and available at PhysioNet.~\cite{Physionet, physionet-web} To
exclude very closely positioned electrodes we used the data from
only 19 electrodes (corresponding to the international 10-20
system).

Each subject was recorded in two different experimental conditions:
one minute baseline recording (rest, eyes open) and total of six
minutes of recordings of fist movement tasks. During the task
period, each subject would perform a series of visually triggered 4
seconds long series of opening and closing fist movements
(followed by 4 second rest intervals excluded from the analysis). In this study, we analyze
three different groups of data: \emph{Baseline All},
\emph{Task All}, and \emph{Task C3-C4}. Baseline All includes data
from all 19 EEG electrodes during baseline recordings.  Task All
includes data from all 19 EEG electrodes during the task periods.
Task C3-C4 includes only the data from C3 and C4 electrodes (which
are close to the motor cortex) during the task periods.

\section{Analysis techniques}
Phase domain is an appropriate way to analyze weakly synchronized
neural signals.~\cite{hurtado_rubchinsky_2004, levanquyen, lachaux,
tass1998} As the coupling strength increases from low to moderate
values synchrony may be observed in the phase domain while the
amplitudes of oscillations remain
uncorrelated.~\cite{arkady_pikovsky} Thus phase may provide a more
sensitive and appropriate metric to explore the relatively
moderately synchronized dynamics we study here.

All signals were referenced to the mean EEG of two ears. EEGs first
filtered in $\alpha (8 - 13 \text{ Hz})$ and $\beta (13 - 30
\text{Hz})$ frequency bands with Kaiser windowed digital FIR filter
sampled at 160 Hz and zero-phase filtering was implemented to avoid
phase distortions. Phase was then extracted via Hilbert transform
resulting in the time-series of phases (see
Refs.~\onlinecite{arkady_pikovsky, hurtado_rubchinsky_2004}). For
each pair of this time-series (measured at the same time)
$\phi_k(t)$ and $\phi_l(t)$ we consider a standard index to
characterize the strength of the phase locking between these two
signals:
$$\gamma = || \frac{1}{N} \sum_{j=1}^N e^{i \theta (t_j)} ||^2,$$ where $\theta (t_j) = \phi_{k}(t_j) - \phi_{l} (t_j)$ is the phase difference.
 This synchronization index varies from 0 (complete lack of phase locking) to 1 (perfect phase locking).
 However, this phase locking index is not designed to describe the fine temporal structure
 of the dynamics, rather it provides an overall index of phase synchrony.~\cite{park_rubchinsky1, ahn_park_rubchinsky}
 Thus even if evaluated on the short time window, it necessarily should include sufficiently large number of
 oscillations and can not be used to inspect if the oscillations are at the preferred phase lag or not at each cycle.

\begin{figure}[htp]
   \centering
   \includegraphics[width=2in,height=2in]{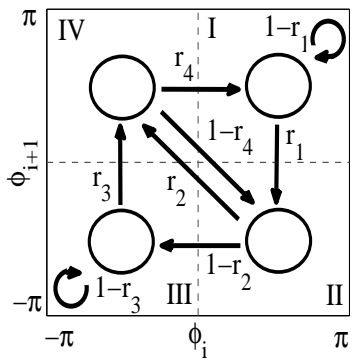}
  \caption{\label{phase_diagram} Diagram of the $(\phi_{i}, \phi_{i+1})$ map. The arrows indicate all possible transitions between regions and the expressions next to the arrows indicate the rates for these transitions. }
\end{figure}

In order to assess the fine temporal structure of synchronous
dynamics of the signal we employed the analysis of phase
synchronization on short time scales via first-return
maps.~\cite{ahn_park_rubchinsky} This method allows for an analysis
of the temporal development of phase difference if some level of
synchrony (some preferred phase-locking angle) is present. Briefly,
whenever the phase of the reference signal crossed from negative to
positive values, we recorded the phase of the other signal,
generating a set of consecutive phase values $\{\phi_i\}_{i=1}^N,$
where $N$ is the number of such crossing. Then $(\phi_i,
\phi_{i+1})_{i=1}^{N-1}$ was plotted. The predominantly synchronous
dynamics appeared as a cluster of points, with the center at the
diagonal $\phi_{i+1} = \phi_{i}.$  We used the Kolmogorov-Smirnov
test to detect non-uniform distribution of $\{\phi_i\}_{i=1}^N$ with
the significance level of 0.05 to include a recording in the further
analysis (the results were not qualitatively affected by this
level). After determining the center of the cluster for each pair of
analyzed signals, all values of the phases were shifted by the same
amount to position the center of the cluster to the center of the
region I (see Fig.~\ref{phase_diagram}).

This phase space was then partitioned into four regions numbered in a clockwise manner, since this is the primary direction of
the dynamics.
The region I is centered around the state with the most frequently observed (preferred) phase difference, and is defined as a synchronized state. In this sense, other three regions are considered as desynchronized states. Note that the system can stay in the third region near diagonal for several iterations of the map (several cycles of oscillations). However, this happens relatively rarely. So region I is the region which corresponds to the most preferred time lag; it is a synchronized state in this essentially data-driven approach.  Thus the
synchronized state here is the one where the deviation from the
preferred phase angle is less than $\pi/2.$

Transition rates $r_{1,2,3,4}$ for the transitions between four
regions of the map are defined as the number of points in a region,
from which the trajectory leaves the region to another region,
divided by the total number of points in the original region (see
Fig.~\ref{phase_diagram}). For example, $r_1$ is the ratio of the
number of trajectories escaping the region I for the region II to the number of all points in
the region I. All rates vary between 0
and 1. The transition rates $r_{2,3,4}$ are related to the
durations of desynchronizations and define them completely if
transitions are independent. However, as we will show below, transitions were not fully independent in the data
considered here. In addition to this four-state model we will also consider a two-state model, where all three desynchronized states are lumped together into a single desynchronized state.

\section{\label{results} Transitions between Synchronized and Desynchronized States and Durations of Desynchronization Intervals}

\begin{figure}
   \centering
   \includegraphics[width=3.2in,height=2.7in]{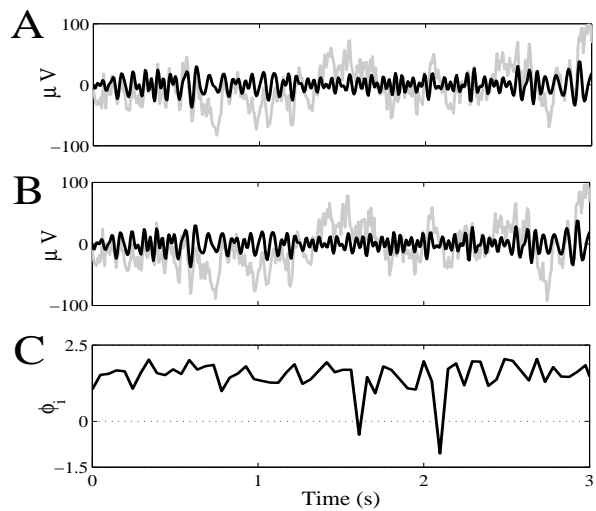}
  \caption{\label{raw_filtered_data} An example of raw (light gray line)
  and filtered EEGs (solid line) for 3 seconds during task execution from C3 (panel A) and C4 (panel B) electrodes
  at the beta band. (C) The phase difference of the above signals plotted as $\phi_i(t_j)$,
  where $t_j$ is the time when the phase of C3 crosses from negative to positive values as described in the text.}
\end{figure}

 Figs.~\ref{raw_filtered_data}A and
~\ref{raw_filtered_data}B show examples of raw data and its
corresponding filtered data for 3 seconds during task execution from
C3 (panel A) and C4 (panel B) electrodes at the beta band.
Fig.~\ref{raw_filtered_data}C shows that the preferred phase
difference between two signals (presented here as
$\phi_i(t_j)$) is bounded (although within relatively
large bounds fitting the discussion in the Analysis Techniques
above) so that two phases are locked for a prolonged (with respect
to the period of oscillations) intervals, interrupted by escapes to
desynchronous states.

\begin{figure}
   \centering
   \includegraphics[width=3.2in,height=2.7in]{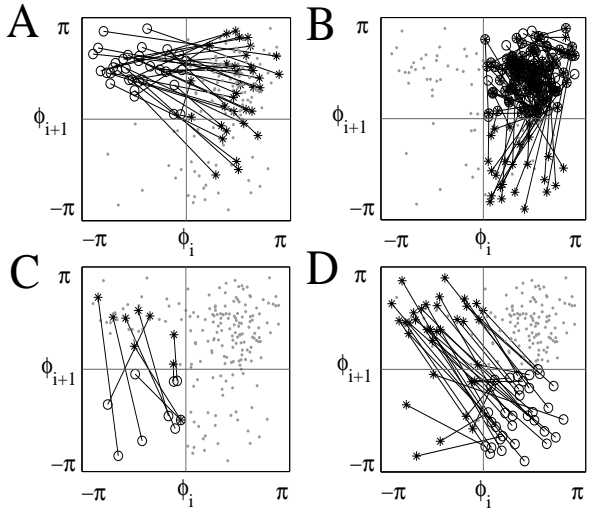}
  \caption{\label{return_map_example} An example of the first-return map for C3 and C4 EEGs for one subject during Baseline eyes open for the first 15 seconds. All four plots have the same data points (gray dots), but each subplot (A-D) presents the evolution of points from one region. If a point evolves from one region to another region, then we represent it as $\circ-\ast$. If a point evolves within the same region, then we represent it as
  $\circ-\circ$.}
\end{figure}

An example of first-return map from the data is shown at
Fig.~\ref{return_map_example}. The transition rates $r_{1,2,3,4}$
correspond to panels (B), (D), (C), and (A), respectively. In one
map iteration, most points in region I evolve into points within the
region while relatively few points evolve to region II
(Fig.~\ref{return_map_example}B). So $r_1$ is relatively small.
Other transition rates are computed similarly.

We also computed the relative frequencies (probabilities) of
desynchronization events of different durations. In the considered
first-return map approach, the duration of a desynchronization event
is the number of steps that system spends away from region I minus
one.  This number of steps minus one is essentially a number of
cycles of oscillations the signals are desynchronized. The shortest
duration of a desynchronization event corresponds to the shortest
path II $\rightarrow$ IV $\rightarrow$ I. This corresponds to the
desynchronization length of one cycle (in other words, in two steps
the phases are back in a locked state).

\begin{figure}[htp]
   \centering
   \includegraphics[width=3.2in,height=2.7in]{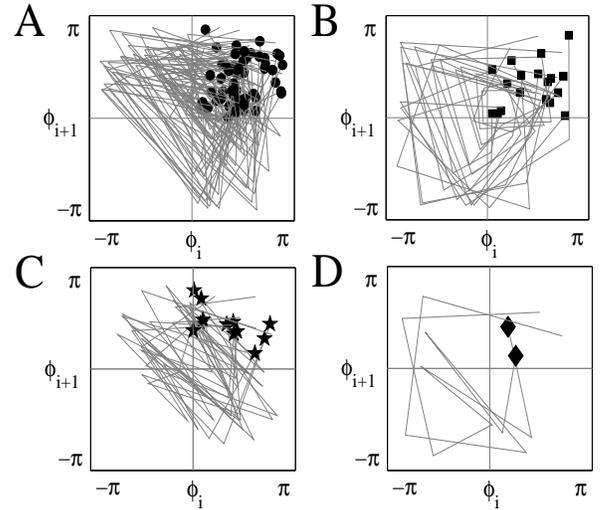}
  \caption{\label{duration_example} Examples of desynchronizations of different
  durations (same data as in Fig.~\ref{return_map_example} for 40
  seconds). The gray line represents the trajectory during a desynchronization episode. For example, the trajectory for the shortest possible desynchronization (panel A)
leaves region I, passes through regions II and IV, and returns to
region I. Bold symbols in each panel represent the initial points
for each desynchronization. (A)-(D) present all desynchronizations
of duration of one to four cycles of oscillations respectively.}
\end{figure}

Fig.~\ref{duration_example} shows examples of desynchronization
events of different durations.  The
number of origination points for the desynchronizations lasting for
one cycle of oscillations (Fig.~\ref{duration_example}A) was much
larger than that for other durations. This suggests that the
probability of the desynchronization lasting for one cycle was high
while probabilities of longer durations were low. We now present the
cumulative results for all 109 healthy human subjects for three
different cases: Baseline All, Task All, and Task C3-C4.

At both $\alpha$ and $\beta$ frequency bands, the means of overall synchrony index
$\gamma$ between any two EEGs were between 0.18 and 0.43 for all
three different cases. That is, the overall levels of synchrony were
moderate. The results for both frequency bands were
largely similar to each other. Thus, we will present the
illustrations only for $\beta$ band.

\begin{figure}[htp]
    \centering
   \includegraphics[width=3.2in,height=2.5in]{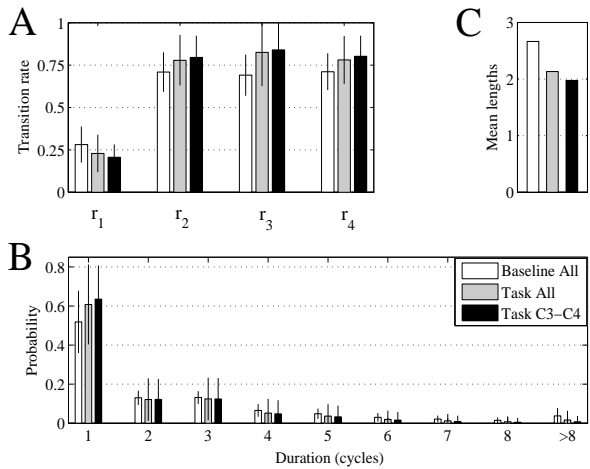}
  \caption{\label{beta_analysis}(A) Transition rates $r_{1,2,3,4}$, (B) distribution of desynchronization durations,
  (C) mean duration of desynchronizations at $\beta$-band.
  The last bin of the histogram (B), ``$>8$", is a sum of the relative frequencies
  of all desynchronizations longer than eight. Mean $\pm$ SD is presented at A and B. }
\end{figure}

The transition rates and distributions of durations of desynchronizations are presented in Fig.~\ref{beta_analysis}.
The rate $r_{1}$ for all three cases (Baseline All, Task All, and
Task C3-C4) was significantly lower than 0.3 while rates
$r_{2,3,4}$ were significantly higher than 0.6 ($p < 10^{-16}$,
t-test was used here and below). These low $r_1$ and high $r_{2,3,4}$
values promote high probability of short desynchronization episodes. Note that in spite of their overall similarity, the
transition rates $r_{1,2,3,4}$ vary across different conditions (baseline and task)
and set of electrodes (all and C3-C4), as should be expected as EEGs
in these cases should reflect different underlying neurophysiology.

Fig.~\ref{beta_analysis}B shows that the probability to observe
desynchronization lasting for one cycle of oscillations was
significantly higher than 0.5 while the probabilities of longer
desynchronizations were significantly lower than 0.2 $(p <
10^{-16})$ for all three cases considered. The probabilities to
observe the shortest desynchronization (length of one
cycle of oscillations)  for all three cases were at least 3 times
higher than the probabilities of other
lengths of desynchronizations (including those lasting for two  or
three cycles $(p < 10^{-16})$). These high probabilities of the
shortest desynchronization (length of one cycle of
oscillations) imply the short mean desynchronization duration.
The mean lengths of desynchronization episodes were between
1.9 and 2.7 (Fig.~\ref{beta_analysis}C). The modes and medians of
distributions of desynchronization durations were always just one
cycle of oscillations. For all pairs of signals analyzed, for
Baseline All only in 0.06\% of cases the desynchronization duration
of one cycle was less frequent than the desynchronization duration
of two or three cycles. For the Task All and Task C3-C4 this
fraction was 4.9\% and 3.4\%.

The rates $r_{1,2,3,4}$, the averages and the distributions of durations of desynchronization events for Task All and Task
C3-C4 were more or less similar to each other and more different from those distributions for Baseline All. This is not surprising because subjects performed a motor task in response to an external cue, which may involve a change in oscillatory activity in large cortical areas.

\begin{figure}[htp]
    \centering
   \includegraphics[width=3.2in,height=2.5in]{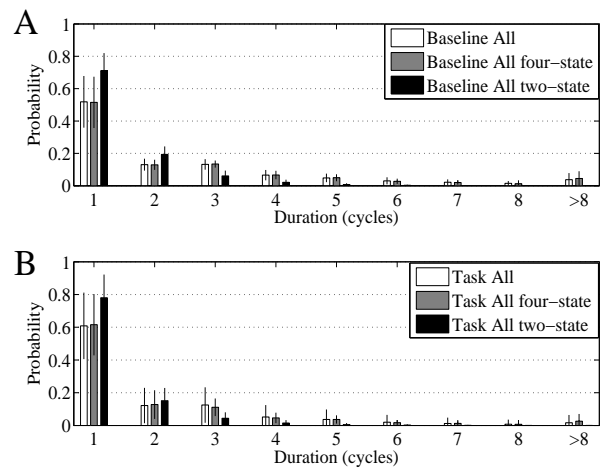}
  \caption{\label{two_four_states} Distribution of desynchronization durations from the data (white bars, the same as in Fig.~\ref{beta_analysis}), from the four-state model rates $r_{1,2,3,4}$(Fig. 1), and two-state model rates. (A) Baseline dynamics. (B) Dynamics during the task performance. }
\end{figure}

We now will consider how well the observed desynchronization
intervals may be described in the framework of a Markov chain model:
independent transitions between synchronized and desynchronized
states. Fig.~\ref{two_four_states} shows the distribution of
desynchronization durations with two-state model and four-state
model. In the two-state model there are synchronized state (region
I) and the desynchronized state (regions II, III, and IV). In this
model, there are only two transition rates: $\hat{r}_1$  (transition rate from the synchronized state to the
desynchronized state) and $r_R$ (return rate of resynchronization).
The duration of desynchronization is the number of time-steps that
system spends in the desynchronized state.

As can be seen at Fig.~\ref{two_four_states}, the distributions of
desynchronization durations generated by four-state model were
almost (although not completely) identical to the distribution of
durations obtained directly from the data. Although the
distributions  generated by the two-state model were visibly
different from those from the data, the general pattern of the
distributions was well captured by two-state model as well. That is,
the probability of the shortest desynchronization
(length of one cycle of oscillations) is much larger
than probabilities for other
desynchronizations.

An interesting observation is that although the rates $r_{1,2,3,4}$ depend on conditions (Baseline or Task) and set of electrodes, $r_1 + r_{2,3,4} \approx 1$(see Fig.~\ref{beta_analysis}A). This leads to the following ramification. The transition matrix of two-state model is
\[  \left( \begin{array}{cc}
1-\hat{r}_1 & \hat{r}_1 \\
r_R & 1-r_R \end{array} \right).\] If $\hat{r}_1 + r_R=1$, then this
system is in the detailed balance. Moreover, the eigenvalues of this matrix are 0 and 1.  So that if the
system is perturbed from a stationary distribution, it
settles back to a stationary distribution after just
one time-step (if there is small, but non-zero eigenvalue, the
convergence is exponential, but very fast).
Roughly speaking, $\hat{r}_1 \approx 1/4$ and $r_R\approx 3/4$ so
that for synchronous on the average episodes we look at about a
quarter of pairs of locations of the studied brain network are in
non-synchronous state and about three quarters are in synchronous
state. If this distribution is perturbed, the system gets back to
the stationary state within just one cycle of oscillations given the
observed transition rates.

Similar considerations are valid for the four-state model as well. The transition matrix of four-state model (see Fig. 1) is
\[ \left( \begin{array}{cccc}
1-r_1 & r_1 & 0 & 0 \\
0 & 0 & 1-r_2 &  r_2 \\
0 & 0 & 1-r_3  &r_3\\
r_4 &1-r_4 &0  &0 \end{array} \right).\]
If $1- r_1 = r_2 = r_3 = r_4$, then the eigenvalues of this matrix are 0 (multiplicity of 3) and 1. The square of this matrix projects any vector to the subspace corresponding to the eigenvalue 1. Thus again, if the stationary distribution is perturbed, the system will be back to the stationary distribution of synchronized and desynchronized states very quickly.

According to our measurements $r_1 + r_{2,3,4} $ is close, but not
identical to 1. This can be either due to the features of the
time-series analysis or due to the nature of the observed system.
However even in the latter case the considered brain networks will
be in a state  where the perturbed system rapidly converges to the
stationary distribution.

\begin{figure}[htp]
    \centering
   \includegraphics[width=3.2in,height=2.5in]{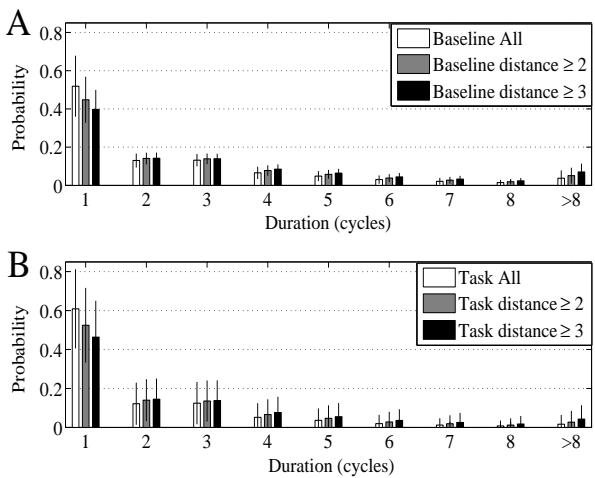}
  \caption{\label{distance23} Distribution of desynchronization durations for all electrode pairs (the same as in Fig.~\ref{beta_analysis}) in comparison with the distributions obtained from pairs of electrodes separated by at least one (distance $\geq 2$), and at least two (distance $\geq 3$) other electrodes. (A) Baseline dynamics. (B) Dynamics during the task performance. }
\end{figure}

Finally, we analyzed the data from pairs of distant electrodes to eliminate volume conduction effects, which may potentially affect the analysis of the patterning of synchrony. We computed the durations of desynchronization events for pairs of electrodes separated by at least one and at least two other electrodes (Fig.~\ref{distance23}). Strictly speaking, electrodes are not exactly equidistant, however, the distance between neighboring electrodes is not very much different and  as we exclude immediate and next neighbor electrode pairs, we obtain results for substantially remote electrodes. While consideration of only remote pairs of electrodes affected the distribution of durations of desynchronization events, this effect was small for both baseline and task activity. In both considered cases the probabilities to observe the shortest possible desynchronizations were at least two times higher than the probabilities of longer desynchronizations  for both Baseline and Task cases $(p < 10^{-16})$.

\section{Discussion}

The present study analyzed the fine temporal structure of neural
synchrony in $\alpha$ and $\beta$ frequency bands in healthy human
EEG during resting state and a motor task. We found that in all
considered cases oscillations go out of synchrony frequently, but
primarily for only a small number of cycles of oscillations. The
chances of longer desynchronization episodes decreased as the
duration of the desynchronization episodes increases. Moreover, the
studied system appears to be in a detailed balance between
synchronized and desynchronized states within the framework of the
two-state model.

Both the two-state model and four-state model can capture the
temporal pattern of the short desynchronization events (length of
one cycle of oscillations) in the considered study. However, the
two-state model cannot distinguish the difference between two
different systems with the same stability of the synchronized state
or synchrony level and $\hat{r}_1$, while the four-state model can
effectively discriminate the difference.~\cite{ahn_park_rubchinsky}
Moreover, from an experimental viewpoint we do not know in a priori
whether the two-state model is enough to capture most of the
important temporal dynamics of complex systems.  We also would like
to reiterate that both models are a way to describe
synchronizations/desynchronizations in pairs of different brain
areas as they develop in time rather than a detection of multiple
patterns of synchronous and nonsynchronous pairs or clusters in
spatially complex partially synchronized regime (like in
Ref.~\onlinecite{rio2011}).

Moderate coupling strength in a system of two coupled oscillators
may induce the same strength of phase-locking, while the dynamics
may be dominated by both long and short desynchronization episodes
depending on the type of oscillators.~\cite{ahn_park_rubchinsky}
Thus different desynchronization patterns provide different means to
generate the same moderate synchrony levels. However, our analysis
of EEG data indicates that the brain networks favor moderate
synchrony with frequent short desynchronizations. While the
particulars of the fine temporal structure in each frequency band
are different, qualitatively our observations hold true both at rest
and during task execution, for pair-wise synchrony across all brain
areas as measured by EEG and for synchrony in the motor cortex. Note
that the analysis of micro-electrode measured neural activity from
subcortical areas of parkinsonian patients also showed the dominance
of short desynchronization events.~\cite{park_rubchinsky1} Thus,
this dominance of short desynchronizations may be a generic feature
of brain networks.

 In line with this, it is interesting to note that
dynamics of the phase synchronization of neuroimaging data from the
human brain shows power law probability distribution, compatible
with dynamical criticality, of both periods of phase-lock interval
and rapid change of synchronization at broad frequency
bands.~\cite{kitzbichler2009} While that study was focused on a
longer time scales (the analysis involved averaging, thus
desynchronization was not followed from cycle to cycle of
oscillations), those results suggest functional implications similar
to ours. This criticality of the human brain may have capacity to
change the configuration rapidly in response to external inputs more
efficiently.~\cite{kitzbichler2011}

Even though the functional significance of the observed distribution
of desynchronization events is yet unknown, we can speculate on
potential functional implications. As noted in the beginning of this
paper, neural synchrony has been conjectured to be important for
several neural functions, including the formation of neuronal
assemblies. Short desynchronizations may be more likely to
facilitate the function of synchrony in the overall low-synchrony
environment, because the synchronous state frequently gets a chance
to reestablish itself (although for short time). Numerous short
desynchronizations (as opposed to infrequent, but long
desynchronizations) are likely to indicate that synchrony is both
easy to form and easy to break. The probability of the shortest
desynchronization (length of one cycle of oscillations) increases
during a task, then it decreases to the baseline state during the
resting state. This may suggests that whenever a cell assembly must
be formed to facilitate a particular function or task, short
desynchronization dynamics may allow for a quick and efficient
formation and break-up of such an assembly. On the other hand, the
ramifications of the transition matrices whose eigenvalues are only
0 and 1 where eigenvalue 1 has multiplicity 1, which we discuss
above, suggest that once the stationary distribution of synchronized
and desynchronized states is perturbed, the system converges back to
this stationary (and presumably beneficial) distribution very fast.

 Finally, we need to note that our analysis does not explore the potential role of noise in our observations. Noise can have a substantial effect on intermittent phenomena, however it is hard to manipulate in neurophysiological experiment. Nevertheless, time-series analysis used here does not make assumptions regarding the noisy component of the data. Moreover, modeling of neural-like (conductance-based) oscillators suggested that transition rates and the duration of desynchronization events are only weakly affected by
noise of mild intensity.~\cite{ahn_park_rubchinsky}

Our results call for a search for dynamical mechanisms responsible for the short desynchronizations.

\begin{acknowledgments}
This study was partially supported by iMMCS-GEIRE (IUPUI) and by NIH R01NS067200 (NSF/NIH CRCNS).
\end{acknowledgments}

\end{document}